\documentclass[twocolumn,showpacs,preprintnumbers]{revtex4}

\usepackage{graphicx} 
\usepackage{dcolumn}
\usepackage{bm}
\usepackage{ifthen}
\usepackage{booktabs}
\usepackage{amsmath}
\usepackage{amssymb}

\usepackage{bm}

\begin{document}

\title{Direct observation of excitonic polaron in InAs/GaAs quantum dots}
\author{Ming Gong}
\affiliation{Key Laboratory of Quantum Information, University of
Science and Technology of China, CAS, Hefei, 230026, People's
Republic of China}
\author{Chuan-Feng Li $\footnote{email: cfli@ustc.edu.cn}$}
\affiliation{Key Laboratory of Quantum Information, University of
Science and Technology of China, CAS, Hefei, 230026, People's
Republic of China}
\author{Geng Chen}
\affiliation{Key Laboratory of Quantum Information, University of
Science and Technology of China, CAS, Hefei, 230026, People's
Republic of China}
\author{Lixin He}
\affiliation{Key Laboratory of Quantum Information, University of
Science and Technology of China, CAS, Hefei, 230026, People's
Republic of China}
\author{Zhi-Chuan Niu}
\affiliation{State Key Laboratory for Superlattices and Microstructures, 
Institute of Semiconductors, CAS, P.O. Box 912, Beijing 100083, People's Republic of China}
\author{F. W. Sun}
\affiliation{Key Laboratory of Quantum Information, University of
Science and Technology of China, CAS, Hefei, 230026, People's
Republic of China}
\author{She-Song Huang}
\affiliation{State Key Laboratory for Superlattices and Microstructures,
Institute of Semiconductors, CAS, P.O. Box 912, Beijing 100083, People's Republic of China}
\author{Yong-Hua Xiong}
\affiliation{State Key Laboratory for Superlattices and Microstructures,
Institute of Semiconductors, CAS, P.O. Box 912, Beijing 100083, People's Republic of China}
\author{Hai-Qiao Ni}
\affiliation{State Key Laboratory for Superlattices and Microstructures,
Institute of Semiconductors, CAS, P.O. Box 912, Beijing 100083, People's Republic of China}
\author{Guang-Can Guo}
\affiliation{Key Laboratory of Quantum Information, University of
Science and Technology of China, CAS, Hefei, 230026, People's
Republic of China}
\date{\today }

\begin{abstract}

   Excitonic polaron is directly demonstrated for the first time
in InAs/GaAs quantum dots with photoluminescence method. A
new  peak ($s'$) below the ground state of exciton ($s$) comes out as
the temperature varies from 4.2 K to 285 K, and a huge anticrossing energy
of 31 meV between $s'$ and $s$ is observed at 225 K, which can only
be explained by the formation of excitonic polaron. The results also provide
a strong evidence for the invalidity 
of Huang-Rhys formulism in dealing with carrier-longitudinal optical phonon
interaction in quantum dot. Instead,  we propose a simple two-band model, 
and it fits the experimental data quite well. The reason for the finding of 
the anticrossing  is also discussed.

\end{abstract}

\pacs{60.20.Kr, 71.35.-y, 71.38.-k, 78.67.Hc}
\maketitle

Electron-phonon interaction is a very important ingredient
determining the physical properties of semiconductors, 
such as  phonon-assisted  hot carrier relaxation 
process\cite{zibik04, inoshita92}, light absorption and  
emission process\cite{vasilevskiy04,
stauber06}. 
For electron-longitudinal optical (LO) interaction in the weak polar 
system (e.g., GaAs, $\alpha_c=0.067$), the well-documented Huang-Rhys 
model always gives good 
explanation in bulk material. However, in quantum dot (QD),
this  interaction is greatly enhanced 
owing to the discrete nature of energy levels  
with spacing comparable to the energy of LO phonon. Both    
theoretical and experimental findings show that it
may  have entered the strong coupling regime\cite{hameau99, 
vasilevskiy04, verzelen02, sauvage02}, which means that an accurate description 
of this interaction system should be the hybridation of electron state 
and the phonon state, thus the polaron as a new ground state will be 
formed\cite{huybrechts77}. Similar conclusion can also be drawn to hole and exciton 
interacting with LO phonons in QD.   

Although  Huang-Rhys parameter $S$ and Fr\"ohlich coupling constant $\alpha_c$ are both
related to the average number of  LO phonons, while the  irreversible
emission of  LO phonons (0.1 $\sim$ 1 ps) provides an efficient channel for energy 
relaxation\cite{xinqi99}, the formation of polaron will suppress  the LO
phonon contribution to the carrier decoherence process and hence leads to 
long polaron lifetime\cite{sauvage02, verzelen02, hameau99} and 
everlasting oscillation of survival probability\cite{hameau99, hameau02,
deleporte02}.

In experiments, far-infrared (FIR) absorption  results have evidenced the formation
 of electron-LO\cite{deleporte02, hameau02},  hole-LO \cite{preisler05} and
exciton-LO\cite{preisler06} induced magneto-polaron in QDs under ultra-high magnetic
field (up to  $\sim 28$ T) at about 4 K, and anticrossing between the polaron 
levels  differing by one LO phonon\cite{verzelen02} is found. 
However, hampered by the large full width at half maximum (FWHM), 
no direct observation of excitonic polaron has yet been reported. 

In this letter, as we vary the temperature from 4 K to 285 K, the FWHM of the $s$ peak
decreases sharply, thus we can directly observe the excitonic polaron without 
applying magnetic field. It is shown 
that at low temperature, the exciton-LO phonon interaction is weak 
and may be explained within the framework of  
Huang-Rhys model and enhanced $S$ value is obtained. But at high 
temperature,  due to the increasing of coupling strength, excitonic polaron is 
formed, and Huang-Rhys model is invalid.  

The self-assembled InAs/GaAs QDs studied here were grown by molecular-beam
epitaxy (MBE) on GaAs (001) substrate at $515^{\circ}$ C. The sheet
density, mean diameter, and height of the dots are  $\sim$85 $\mu m^{-2}$,
39 nm and  2.8 nm, respectively. The sample is mounted in a He cooled
cryostat and the temperature is  tuned from 4.2 K to 285 K, in the spacing of 7.5 K.
Photoluminescence (PL) was performed 
with an He-Ne laser and a spectrometer with
focal length of 0.75 m equipped with InGaAs line array. The
excitation power varies from 5 mW to 0.5 $\mu$W.

\begin{figure}
\begin{center}
\includegraphics[width=3.0in]{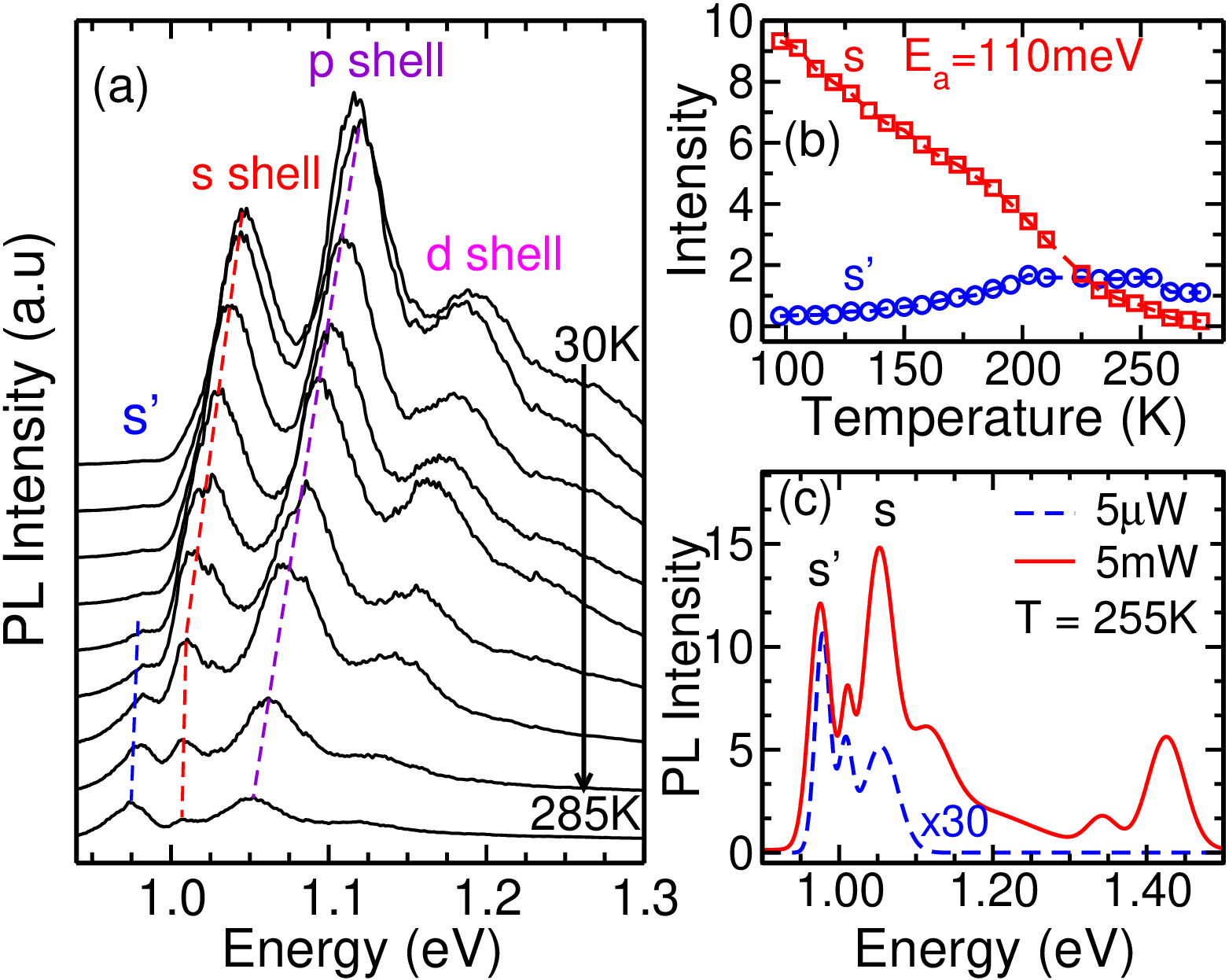}
\end{center}
\caption{(Color online). (a) The  PL spectra of InAs/GaAs QDs ensemble
with respect to temperature from 4.2 K to 285 K at pump-power 5mW, the 
dashed lines  are just for guide.  The
curves have been offset for clarity. (b) The integrated
intensity of $s'$ and $s$ as a function of temperature, $E_a$ is  the
activation energy of $s$ shell.
 (c) The PL spectra at 255 K for different pump-power, the 5 $\mu$W
PL results have been multiplied by a factor of 30.}
\label{PL2temperature}
\end{figure}

 Figure \ref{PL2temperature}(a) presents the PL results as a function of
temperature from  4.2 K to 285 K at pump-power of 5 mW.
$s$, $p$ and $d$  shells, origin of which 
have been well studied by Bayer {\it et al}\cite{bayer00b}, are also the 
zero-phonon line. Strikingly, there is a new peak ($s'$) on the lower-energy 
side of  $s$ shell with quite  different behavior. When T $<60$ K,  
this peak is invisible  even pumped with high  power. 
As we increase the lattice temperature, this peak appears more
and more clear, but the  $s$ peak is  greatly 
suppressed. The intensity of $s$ peak decreases monotonically, whereas,
the intensity of $s'$ increases when  T $<$ 225 K, and then decreases a little
from 225 K to 285 K as shown in Fig. \ref{PL2temperature} (b).
The activation energy $E_a$ of the thermal quenching process for $s$, $p$ and
$d$ shells are 110, 80, 56 meV, respectively, when fitting with 
$I(T)=I_0/(1+A\exp(-E_a/T))$\cite{leymarie95}. 
Similar peaks have also been found in single quantum dot by Bayer {\it et al},
that arise from "Hidden symmetry" effect\cite{bayer00b}, but as we reduce
the pump-power from 5 mW to sufficient low conditions shown in Fig. 
\ref{PL2temperature} (c), the $s'$ still exist, implying that this new peak
can not be accounted for this effect. It also can not from 
multicharge-exciton effect\cite{hartmann00, heitz00}, because it can not 
introduce so large redshift.

\begin{figure}
\begin{center}
\includegraphics[width=3in]{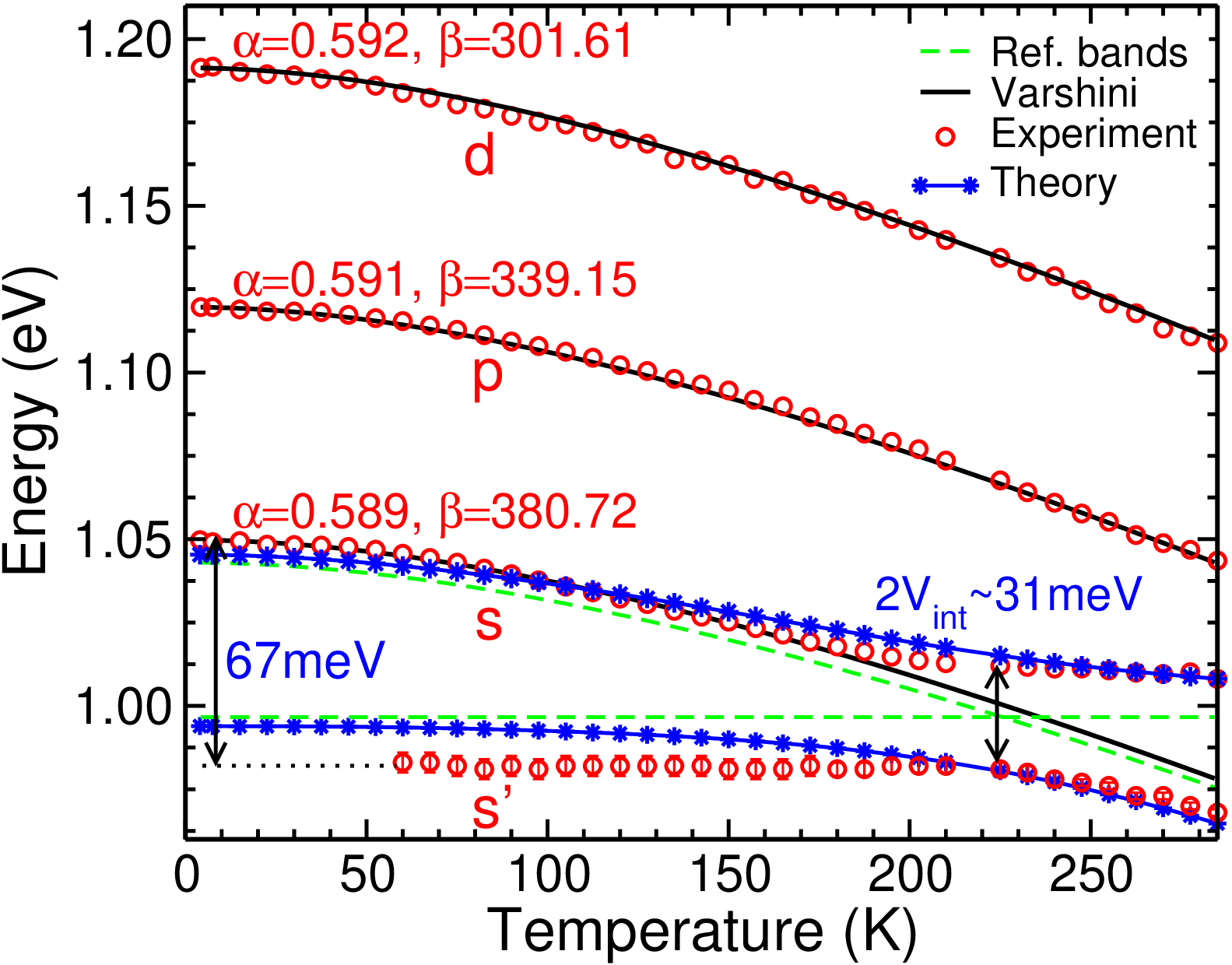}
\end{center}
\caption{(Color online). The emission lines with respect to 
temperature at pump-power 5mW. The open circle is the experimental 
results for $s'$,
$s$, $p$ and $d$, while the solid line is the results fitted with
Equation (\ref{eq:vershni}), $\alpha$ and
$\beta$ are the corresponding fitting results. The star is the
fitting results with our two-band model, 
while  the dashed line is the corresponding reference bands (see the text). 
The huge anticrossing energy is about 31meV at 225K.} 
\label{fitresults}
\end{figure}

The peak energy positions of $s'$, $s$, $p$ and $d$  
are presented in Figure \ref{fitresults} with open cicle.  Due to the
scattering by phonons\cite{favero03, besombes01, fan51}, $s$,
$p$, $d$ peaks redshift with increasing temperature, however, the
$s'$ peak does not redshift  from 60 K to 170 K. When T $>$ 170 K, it begins to
redshift, but the $s$ shell seems to be  "frozen" at about 1.01 eV, which
is quite similar to the theoretical results in the intermediate and strong
coupling regimes ($\alpha_c > 3$)\cite{huybrechts77}.
From 225 K to 285 K, $s$ shell only redshifts less than 4 meV while the
$p$, $d$ shells redshift more than 25 meV (see below). 

Generally, the peak position  as a function of temperature can be 
 well fitted with Varshni formula in bulk material, and even in single 
QD when T $<$ 100 K\cite{ortner05a}. The empirical Varshni formula is

  \begin{equation}
  E(T)=E(0)-\frac{\alpha T^2}{\beta+T},
  \label{eq:vershni}
  \end{equation}
where $\alpha$  and $\beta$  are fitting parameters characteristic
of a given material.
The solid line in Fig. \ref{fitresults} is the fitting
results with this equation  
for $s$, $p$, and $d$  peaks.
For $p$ and $d$ peaks, this formula
fits the results quite well with error less than 2 meV through the whole
experiment range. For $s$ peak, it fits  well  when T $<$170 K, while in
the high temperature regime, huge deviation is found.  Interestingly,
for these three peaks, we find that $\alpha$ are almost the same
($\sim$0.590 meV/K), and for  $\beta$,  $\beta_s(387.5~\text{K}) >
\beta_p(339.2~\text{K}) >\beta_d(301.6~\text{K})$, these values are larger than
those in bulk material due to the confinement effect 
($\beta=93$ K for InAs\cite{vurgaftmana01}). $(\alpha/\beta)_s=1.55\times
10^{-3}$ is also quite close to the results in  single
QD\cite{ortner05a}.  

In  Fig. \ref{fitresults}, an anticrossing between $s$ and $s'$, with 
crossing energy about 31 meV, is 
found at 225 K, which indicates that the $s'$ peak can not be 
accounted for emission of one LO phonon from $s$\cite{minnaert01,
bissiri00,heitz99}; and thus it is believed to be
a direct evidence for the formation 
of polaron. To understand this result, we propose a two-band 
model following the spirit of Preisler {\it et al}\cite{preisler06}.
We assume the two reference bands $|s\rangle$ and $|s'\rangle$ satisfy
(i) The energy of $|s\rangle$ can be described by Varshni
formula. (ii) The energy of $|s'\rangle$ is independent of
temperature. Under the strong coupling conditions,
the polaron state should be the entanglement of exciton levels  and phonons,
thus $|s'\rangle=\sum_{i,j,{n_{\bf q}}}c_{i,j,\{n_{\bf q}\}}|
e_ih_j \{n_{\bf q}\} \rangle$
\cite{verzelen02}, where ${\bf q}$ is the different modes of the LO phonon,
$e_i$ and $h_j$ represent the electron and hole energy levels, respectively.
We adopt the same  Fr\"ohlich  Hamitonian as Refs.
[\onlinecite{preisler06, verzelen02}], so the off-diagonal matrix element can
be read as $\langle s|(H_F^e+H_F^h)|s' \rangle= A_F
\sum_{i,j,{n_q}}c_{i,j,\{n_q\}} q^{-1} (\sqrt{n_q}\langle s|(e^{-i{\bf
q}\cdot {\bf r}_e}-e^{-i{\bf q}\cdot {\bf r}_h})|e_ih_j \{n_q-1\} \rangle
+ \sqrt{n_q+1} \langle s|(e^{i{\bf q}\cdot {\bf r}_e}-e^{i{\bf
q}\cdot {\bf r}_h})|e_ih_j \{n_q+1\} \rangle)$,  
where $A_F \propto \sqrt{\alpha_c}$\cite{huybrechts77}, 
$n_{\bf q}=(e^{\hbar\omega_q/T}-1)^{-1}$ is the number of LO phonons. The 
nonvanishing coupling only occurs between the factorized 
exciton-phonon states which differ 
by one LO phonon.
The acoustic phonon is neglected for the coupling strength to exciton is greatly 
suppressed in QD\cite{bockelmann90, hui02}, while the effect of lattice 
vibrations on the energy gap is included in $E_s$\cite{fan51}.
Furthermore, dispersionless LO phonon approximation is used, then 
$n_{\bf q}$ can be replaced by the average number of 
LO phonon $n_0$, and the matrix element can be simplified to 
$\langle s|(H_F^e+H_F^h)|s' \rangle= \sqrt{n_0}v_{-}
+\sqrt{n_0+1}v_{+}$, where $v_{\pm}=A_F \sum_{i,j,{n_q}}c_{i,j,\{n_q\}}
q^{-1} \langle s|(e^{\pm i{\bf q}\cdot {\bf r}_e}-e^{\pm i{\bf q}\cdot {\bf
r}_h})|e_ih_j \{n_q\pm 1\} \rangle$. 
For simplicity, we assume that $v_{-}=v_{+}=V_0$,  then the 
off-diagonal element can be reduced to $V_{\text{int}}(T)=
(\sqrt{n_0}+\sqrt{n_0+1})V_0$. The total Hamitonian to
describe this anticrossing  gives

   \begin{gather}
    H=
    \begin{bmatrix} E_s(T)      & (\sqrt{n_0}+\sqrt{n_0+1})V_0          \\
                    (\sqrt{n_0}+\sqrt{n_0+1})V_0^*          & E_{s'}     \\
    \end{bmatrix},
    \label{eq:hamitonian}
    \end{gather}
where $E_{s'}$ is chosen to be 
 $(E_{\text{polaron}}(225 ~\text{K})+ E_{\text{exciton}}(225 ~\text{K}))/2=
0.996$ eV, $E_s(T)=E_s(0)-\alpha T^2/(\beta+T)$,
$\alpha=0.589$ meV/K is the same as the previous fitting results,
and $\beta$ serves as the only fitting parameter here.  From Equation
(\ref{eq:hamitonian}), we have the anticrossing energy 
to be $2(\sqrt{n_0}+\sqrt{n_0+1})|V_0|=31$ meV.
The LO phonon energy we choose here is 
$\hbar\omega_{\text{LO}}=36$ meV\cite{preisler06,minnaert01,hameau99} 
(bulk GaAs-like), so we have $|V_0|=10.1$ meV,
which is slightly larger than that  obtained  under high magnetic 
field\cite{preisler06,hameau02,verzelen02}.  The small discrepancy 
 may be accounted for  the Coulomb interaction between exciton and 
polaron, which is not included in Eq. (\ref{eq:hamitonian}). 
$\beta=412$ K is our fitting result. 

$E_s$ and $E_{s'}$ are  shown with dashed lines in Fig. \ref{fitresults}, 
while the fitting results  with
this two-band model are presented by star. We can 
see that our theoretical calculation fits the 
 experimental results quite well.  For the $s$ shell, it has recovered
almost all the experimental results from 4 K to 285 K, especially at
high temperature, perfectly reproducing the "frozen" state. 
For $s'$ state, it fits the results well
from 185 K to 285 K; however, at low temperature, the
deviation is about 10 meV, because of the simple model we used.
Results show that this two-band model only work well near 
the resonant condition. 

 It is worthy to underline 
that the position of this anticrossing  differs
greatly from that found with FIR  magnetospectroscopy method\cite{preisler06}.
In our results,  the anticrossing is between the exciton state ($s$) 
and the polaron state ($s'$), but not between the polaron levels (intra-band) that 
differ with one LO phonon\cite{preisler06, verzelen02}. It also implies 
that the polaron and exciton may coexist in a  single QD, and the Coulomb 
interaction between them may play a role in the 
formation of polaron (see below).

Note that at 4.2 K, the level spacing of
$s'$ and $s$ is  $67$ meV, about twice the energy of  
$\hbar\omega_{\text{LO}}$\cite{minnaert01,hameau99}.
Actually, phonon replicas at low temperature 
in the PL spectra have been  found and
explained by enhanced $S$ values\cite{odnoblyudov99, heitz99}. 
In this model, their integrated intensity is related to 
the transition  probability $w_p=|\int dQ \chi_p^*(Q-Q_0)\chi_0(Q)|^2=
S^p e^{-S}/p!$, and the ratio between term $p$ and term $(p-1)$ 
gives $S/p$.  At low temperature, 
it is  reasonable to assume that the $s'$ is
originated from exciton recombination by emitting 2LO 
phonons\cite{bissiri00, minnaert01}, then we 
get $S=\sqrt{2I_{s'}/I_s}\leq 0.2$ (in
 bulk InAs, $S_{\text{InAs}}\sim 0.0033$\cite{heitz99}), in consistent 
with  the experimental results in Table I of Ref.
[\onlinecite{bissiri00}]. However, because of the 
 large inhomogeneous broadening due to fluctuation in the dot shape 
and/or size ($\text{FWHM}=32\sim 37$ meV), 
it is hard to separate the -LO satelite line from 
the $s$ shell at low temperature\cite{bissiri00}.

\begin{figure}
\begin{center}
\includegraphics[width=3in]{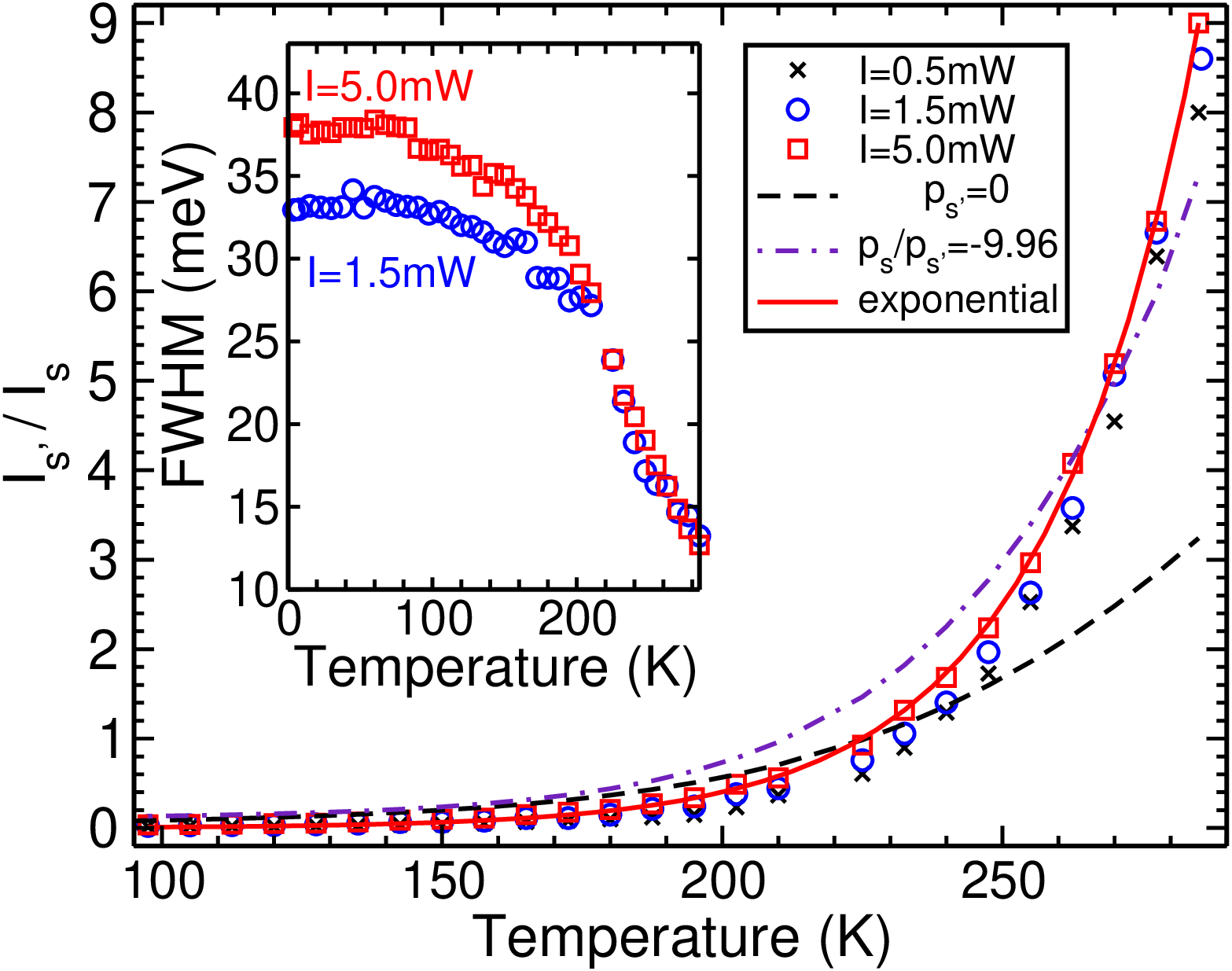}
\end{center}
\caption{(Color online). $\text{I}_{s'}/\text{I}_s$ with respect 
to temperature for
different pump-power (I=0.5, 1.5 and 5 mW), the dashed line is
the fitted results with our two-band model in the dark state limit ($p_{s'}=0$),
 the dash dotted line is the fitted result assuming
$p_s/p_{s'}=-9.96$, the solid line is the fitted results with $A
e^{\alpha_e T}$ for I=5 mW. 
The inset is the FWHM of $s$ shell as  a  function of
temperature for I=1.5 and 5 mW.} 
\label{slotos}
\end{figure}

  The FWHM of $s$ shell with respect to temperature  is presented in
the inset of Figure \ref{slotos}. 
Different from single QD, the FWHM of which increases
monotonically  with the  increasing of temperature\cite{bayer02, hui02},  
the FWHM of $s$ shell decreases from
37 meV at 4.2 K to $\sim$12 meV at 285 K. Many body scattering induced
FWHM broadening is also observed\cite{heitz00}, but quite small when
$\text{T}>200$ K. Similar
anomalous decrease of FWHM is also found in quantum well due to 
the exciton thermalization effect\cite{xu96}.  This unexpected 
result is also one of
the advantage in our experiment. The FWHM of $s'$ peak is
less than 13 meV at 285 K, which is much smaller than the spacing 
between $s$ and $s'$,  so the emission line of $s$ and $s'$ can be separated  and 
identified from PL spectra unambiguously.

Following the method used in Refs. [\onlinecite{bissiri00, odnoblyudov99}], 
we calculate the ratio of intensity of $s'$ to $s$. 
The experimental results are presented in Fig.
\ref{slotos} for different pump-power ($\text{I}=0.5$, 1.5 and 5 mW).
 It is shown that this value increases almost exponentially with respect to 
temperature, as shown in  Fig. \ref{slotos}, in which  the solid
line is our fitting result with $A\exp(\alpha_e T)$ for $\text{I}=5$ mW.  
For different
excitation power, $\alpha_e$ varies from 0.041 to 0.045 $\text{K}^{-1}$. On the other
hand, as we increase the pump-power,
$\text{I}_{s'}/\text{I}_{s}$ increases slightly, indicating that the
Coulomb interaction between exciton and polaron plays a positive 
role in the formation of polaron.
This ratio can exceed 8.0 at 285 K, which can not be explained
by Huang-Rhys model and thus supports the argument
that in QDs, exciton-LO phonon interaction  has entered the strong coupling
regime although the system is electrically neutral\cite{verzelen02}. 

 To illustrate the essential physics of the polaron effect clearer,
we calculate the oscillator
strength (OS) of the given polaron state, which has mixed the ingredient of
$|s\rangle$ and $|s'\rangle$. The OS from Fermi's golden rule read as

 \begin{equation}
 O_{\pm} \propto |\langle \Phi_f(T)|H'|\Phi_{\pm} \rangle|^2=
 |\alpha_{\pm} p_s+\beta_{\pm}p_{s'}|^2,
 \end{equation}
where $H'$ is the Hamitonian of laser field  interacting with exciton,
$|\Phi_{\pm} \rangle=\alpha_{\pm} |s\rangle+\beta_{\pm} |s'\rangle$ is
the eigenstate of Eq. (\ref{eq:hamitonian}), 
$p_s= \langle \Phi_f(T)|H'|s\rangle$, and $p_{s'}= \langle
\Phi_f(T)|H'|s'\rangle$. $|\Phi_f(T)\rangle$ is the final state
depending on  the temperature. When $\text{T}\rightarrow 0$, the allowed transitons
are the components containing zero phonon, 
so $|p_s| \gg |p_{s'}|$, and this can 
explain why $s'$ is hard to be observed at low temperature 
(Fig. \ref{PL2temperature} (a)). 
In the dark state limit, we can choose $p_{s'}=0$, then
$I_{s'}/I_s=O_{-}/O_{+}=|\alpha_{-}/\alpha_+|^2$. 
The fitting results under this limit is presented in  Fig. \ref{slotos} with dashed line.
At low temperature, it works quite well, while at 285 K, 
$I_{s'}/I_s \sim 3$ is found. Although this value is  smaller
than the experimental results, the tendency agrees with the
experimental results quite well and the transformation from exciton to
polaron can still be indicated clearly. In fact, 
$p_{s'}$ do not close to zero, for the lack of knowledge about this
quantity,  we further present our fitting result with fixed
$p_s/p_{s'}=-9.96$ through the whole experimental range with dash dotted
line in  Fig. \ref{slotos}, and it gives good fitting 
 especially at high temperature. At 285 K, the OS of 
$|\Phi_-\rangle$ is much stronger than $|\Phi_+\rangle$ ($I_{s'} \gg I_s$), 
which is also responsible for  the suppression of $s$ peak shown
in Fig. \ref{PL2temperature} (a). More importantly, this result 
is also a significant evidence
that the anticrossing found in Fig. \ref{fitresults} is not artificial, but
indeed exists. 

 To sum up, we have demonstrated  the formation of
excitonic polaron directly from  PL results.  A new peak ($s'$) is
found below the $s$ shell, and
 the optical properties of this state  with respect
to temperature are studied in detail. The origin of $s'$ is
not from emission of one LO phonon and/or many body effect.  We propose a simple
two-band model, and it fits both
the peak positions and the ratio of integrated intensity of $s'$ to
$s$ quite well. The reason for the  finding of the anticrossing is also discussed.

In QDs, the enhancement of $S$ up to $\sim 0.02$ can be explained by 
8-band ${\bf k}\cdot{\bf p}$ model taking the  piezoelectric effect 
into account\cite{heitz99}. But 
this  estimation is still one or two orders of magnitude  smaller than
the experimental results\cite{bissiri00}. Other ways to achieve large $S$ values
may include exact diagonalization method\cite{vasilevskiy04} and strong 
Fr\"ohlich coupling effect\cite{minnaert01}. In this letter, the formation
of polaron  provides an unambiguous evidence for the invalidity of Huang-Rhys formulism 
 (adiabatic theory) in dealing with exciton-LO phonon interaction in QD.  Recently, T. J. Devreese 
{\it et al}\cite{devreese07, fomin98} 
emphasize that the effect of non-adiabaticity 
should be taken into account to interpret the surprisingly enhancement of the phonon 
replicas in the PL spectra,  and it is also  expected that this method can pave the way 
for a deeper understanding of our results. In the 
strong coupling regime (e.g., high temperature), we also believe that our results are
important to the understanding of the carrier decoherence and relaxation process
in QD\cite{zibik04, inoshita92, vasilevskiy04, stauber06}.

This work was supported by National Fundamental Research Program, the Innovation
funds from Chinese Academy of Sciences, National Natural Science Foundation of
China (Grant No.60121503 and 60625405) and Chinese Academy of Sciences International
Partnership Project.

\bibliographystyle{apsrev}  
\bibliography{polaron}

\clearpage
\newpage

\end{document}